\documentclass[runningheads,a4paper,uplatex]{article}
\usepackage{amsmath,mathtools,amsthm} 
\usepackage{tikz}
\usetikzlibrary{automata,arrows}
\usepackage{algorithm}
\usepackage{algorithmic}
\usepackage{hyperref}
\usepackage{breakurl}
\usepackage{authblk}

\usepackage{graphicx}
\usepackage{epstopdf}
\usepackage[ruled,linesnumbered,algo2e,vlined]{algorithm2e}

\usepackage{amssymb}

\def\newblock{\hskip .11em plus .33em minus .07em}

\newcommand{\lranggle}[1]{\lrangle{#1}}
\newcommand{\lrangle}[1]{\langle #1 \rangle}
\newcommand{\lrsquare}[1]{[ #1 ]}
\newcommand{\mtt}[1]{\mathtt{#1}}
\newcommand{\mrm}[1]{\mathrm{#1}}
\newcommand{\rst}[2]{#1{\upharpoonright}#2}
\newcommand{\dl}{\texttt{\$}}

\theoremstyle{definition}
\newtheorem{example}{Example}

\title{Enumerating Cryptarithms Using \\Deterministic Finite Automata\footnote{This manuscript comprehends and improves our paper~\cite{NozakiHYS18} in the proceedings of CIAA 2018.}}

\author[1]{Yuki~Nozaki\footnote{He is currently working in VOYAGE GROUP, Inc.}}
\author[1]{Diptarama~Hendrian}
\author[1]{Ryo~Yoshinaka}
\author[2]{Takashi~Horiyama}
\author[1]{Ayumi~Shinohara}

\affil[1]{
  Graduate School of Information Sciences, Tohoku University \newline
  \texttt{yuki\_nozaki@shino.ecei.tohoku.ac.jp}
  \newline
  \texttt{\{diptarama@, ryoshinaka@, ayumis@\}tohoku.ac.jp} 
}
\affil[2]{
  Graduate School of Science and Engineering, Saitama University \newline
  \texttt{horiyama@al.ics.saitama-u.ac.jp} 
}

\begin{document}

\maketitle

\begin{abstract}
A cryptarithm is a mathematical puzzle where given an arithmetic equation written with letters rather than numerals, a player must discover an assignment of numerals on letters that makes the equation hold true.
In this paper, we propose a method to construct a DFA that accepts cryptarithms that admit (unique) solutions for each base.
We implemented the method and constructed a DFA for bases $k \le 7$.
Those DFAs can be used as complete catalogues of cryptarithms,
 whose applications include enumeration of and counting the exact numbers $G_k(n)$ of cryptarithm instances with $n$ digits that admit base-$k$ solutions.
Moreover, explicit formulas for $G_2(n)$ and $G_3(n)$ are given.

\end{abstract}

\section{Introduction}
A cryptarithm is a mathematical puzzle where a given arithmetic formula consisting of letters rather than numerals, players try to find an injective substitution of numerals for letters that makes the formula hold true.
%
Figure~\ref{fig:example} shows a well-known example of a cryptarithm and its solution.
To solve a cryptarithm is, in principle, not quite hard.
One can find a solution (if any) by trying at most $10!$ assignments of numerals on letters, i.e., cryptarithms are solvable by brute force in linear time.
Nevertheless, cryptarithms have been an interesting topic of computer science~\cite{Knuth} and different methods for solving cryptarithms have been proposed~\cite{AbbasianM2009,Luoma2016} including a number of online solvers on the web (e.g.,~\cite{Collins,dCode,Tamura}). 
In fact, although cryptarithms can be solved in linear time under the decimal system,
Eppstein~\cite{ref:epstein1987np} showed that 
to decide whether a given cryptarithm has a solution under the base-$k$ system is strongly NP-complete when $k$ is not fixed.
His discussions involve only arithmetic formulas with just one addition, like the one in Figure~\ref{fig:example}.
Following Eppstein, this paper focuses on such formulas only.
A cryptarithm example that has a binary solution but no decimal solution is shown in Figure~\ref{fig:fukumen_k2}.
\begin{figure}[h]
	\begin{tabular}{c}
	\begin{minipage}{0.46\hsize}
		\begin{center}
			\begin{tabular}{rrrrr}
				& {\tt s} & {\tt e} & {\tt n} & {\tt d} \\
				+ & {\tt m} & {\tt o} & {\tt r} & {\tt e} \\ \hline
				{\tt m} & {\tt o} & {\tt n} & {\tt e} & {\tt y}
			\end{tabular}
	\\[10pt]
			\begin{tabular}{rrrrr}
				& 9 & 5 & 6 & 7 \\
				+ & 1 & 0 & 8 & 5 \\ \hline
				1 & 0 & 6 & 5 & 2
			\end{tabular}
	\caption{Example of a cryptarithm and its solution~\cite{ref:dudeney1924}}
	\label{fig:example}	
		\end{center}
	\end{minipage}
	\hspace{0.01\hsize}
	\begin{minipage}{0.46\hsize}
		\begin{center}
			\begin{tabular}{cc}
				& {\tt P} \\
				+ & {\tt P} \\ \hline
				{\tt P} & {\tt A}
			\end{tabular}
	\\[10pt]
			\begin{tabular}{cc}
				& 1 \\
				+ & 1 \\ \hline
				1 & 0 
			\end{tabular}
	\caption{Cryptarithm solvable under the binary system}
	\label{fig:fukumen_k2}	
		\end{center}
	\end{minipage}
	\end{tabular}
\end{figure}


Our goal is not only to provide a cryptarithm solver but to propose a method to enumerate cryptarithms for different base systems.
Towards the same goal, Endoh et al.~\cite{endo11gpw} presented a method for constructing a deterministic finite automaton (DFA) that accepts cryptarithms solvable under the base-$k$ system for $k =2,3,4$.
Their method constructs the goal DFA as the product of several auxiliary DFAs corresponding to different conditions that solvable cryptarithms must satisfy.
On the other hand, our proposed method constructs the objective DFA directly.
This approach enabled us to construct the goal DFAs for $k \leq 7$.

Those DFAs can be seen as complete catalogues of cryptarithms for different bases.
Once the cryptarithm DFA for base-$k$ arithmetics is constructed, this can be used as a cryptarithm solver using the information added to its states that runs in linear time in the size of the input with no huge coefficient.
Moreover, different types of analyses on cryptarithms are possible with standard techniques on edge-labeled graphs.
For example, one can enumerate all the solvable cryptarithms one by one in the length-lexicographic order.
It is also possible to compute the $m^\text{th}$ solvable cryptarithm quickly without enumerating the first $m-1$ cryptarithms.
Counting the number of solvable cryptarithms of $n$ digits is also easy.
In particular, we derived explicit formulas for the number $G_{k}(n)$ of cryptarithms of $n$ digits solvable under the base-$k$ system for $k=2,3$ as
$G_2(n) = 6\times4^{n-2}-3\times2^{n-2}$ and 
$G_3(n) = 4\times9^{n-1}-2\times5^{n-1}-3^{n-1}$, respectively.


\section{Preliminaries}
For an alphabet $\Sigma$, $\Sigma^*$ and $\Sigma^+$ denote the sets of strings and non-empty strings, respectively.
For a map $\theta$ from an alphabet $\Sigma$ to another $\Delta$, its homomorphic extension from $\Sigma^*$ to $\Delta^*$ is denoted by $\hat{\theta}$.
For a string or a tuple of strings $w$ over $\Delta$, $\rst{\Sigma}{w}$ denotes the subset of $\Sigma$ consisting of letters occurring in $w$.
An \emph{extension} of a function $f\colon A \to B$ is a function $g \colon A' \to B'$ such that $A \subseteq A'$ and $g(x)=f(x)$ for all $x \in A$.
The cardinality of a set $A$ is denoted by $|A|$.
The length of a string $w$ is also denoted by $|w|$.
We let $N_k$ denote the alphabet of numerals $0,\dots,k-1$.

\subsection{Cryptarithms}
A \emph{cryptarithm} is a triple $\vec{w}= \lrangle{w_1,w_2,w_3}$ of non-empty strings over an alphabet $\Sigma$.
Each $w_i$ is called the \emph{$i^{\text{th}}$ term}.
The \emph{size} of $\vec{w}$ is defined to be $\max\{|w_1|,|w_2|,|w_3|\}$.
Any injection from $\rst{\Sigma}{\vec{w}}$ to $N_k$ is called a \emph{base-$k$ assignment} for $\vec{w}$.
Moreover it is a \emph{base-$k$ solution} if it makes the equation $\hat{\theta}(w_1)+\hat{\theta}(w_2)=\hat{\theta}(w_3)$ true when interpreting strings over $N_k$ as numerals in the base-$k$ system:
that is, for $w_{i} = w_{i,|w_i|} \dots w_{i,1}$ with $w_{i,j} \in \Sigma$,
 it holds $\sum_{j=1}^{|w_1|} \theta(w_{1,j}) k^{j-1} + \sum_{j=1}^{|w_2|} \theta(w_{2,j}) k^{j-1} = \sum_{j=1}^{|w_3|} \theta(w_{3,j}) k^{j-1}$
 and $\theta(w_{i,|w_i|}) \neq 0$ for each $i=1,2,3$.
A cryptarithm that admits a solution is said to be \emph{base-$k$ solvable}.

Following Endoh et al.~\cite{endo11gpw},
in order for DFAs to treat cryptarithms, we convert cryptarithms into single strings over $\Sigma \cup \{\dl\}$ with $\dl \notin \Sigma$ by
\[
	\psi(\lrangle{w_1,w_2,w_3}) = w_{1,1}w_{2,1}w_{3,1}\,  w_{1,2}w_{2,2}w_{3,2}\,  \dots w_{1,n}w_{2,n}w_{3,n}\, \dl\dl\dl 
\]
where $w_{i} = w_{i,|w_i|} \dots w_{i,1}$, $n=\max\{|w_1|,|w_2|,|w_3|\}$, and $w_{i,j}=\dl$ for $|w_i| < j \le n$.
Such a string $\psi(\vec{w})$ is called a \emph{cryptarithm sequence}.

\begin{example}\label{ex:sendmoremoney}
Let $\vec{w} = \lrangle{\mtt{send},\mtt{more},\mtt{money}}$.
This admits a unique base-$10$ solution $\theta = \{ {\tt d} \mapsto 7, {\tt e} \mapsto 5,{\tt y} \mapsto 2,{\tt n} \mapsto 6,{\tt r} \mapsto 8,{\tt o} \mapsto 0,{\tt s} \mapsto 9, {\tt m} \mapsto 1 \}$.
The sequential form of $\vec{w}$ is $\psi(\vec{w})=\mtt{dey}\,\mtt{nre}\,\mtt{eon}\,\mtt{smo}\,\mtt{\dl\dl m}\,\mtt{\dl\dl\dl}$.\footnote{For readability a small space is inserted in every three letters.}
\end{example}
We say that two instances $\vec{w}$ and $\vec{v}$ are \emph{equivalent} if there is a bijection $\gamma$ from $\rst{\Sigma}{\vec{w}}$ to $\rst{\Sigma}{\vec{v}}$ such that $\hat{\gamma}(\vec{w}) = \vec{v}$.
In such a case, an injection $\theta \colon \rst{\Sigma}{\vec{v}} \to N_k$ is a base-$k$ solution for $\vec{v}$ if and only if so is $\theta \circ \gamma$ for $\vec{w}$.
Fixing the alphabet to be $\Sigma_k = \{\mtt{a}_1,\dots,\mtt{a}_k\}$, we define the canonical form among equivalent instances.
A base-$k$ cryptarithm $\vec{w} \in (\Sigma_k^*)^3$ is said to be \emph{canonical} if
\begin{itemize}
\item wherever $\mtt{a}_{i+1}$ occurs in the sequential form $\psi(\vec{w})$ of $\vec{w}$, it is after the first occurrence of $\mtt{a}_{i}$ for any $i \ge 1$.
\end{itemize}
Identifying a cryptarithm and its sequential form, we adapt terminology on cryptarithms for cryptarithm sequences as well.
For example, the sequential form of a canonical cryptarithm is also called canonical.
A solution of a cryptarithm instance is also said to be a solution of its sequential form.

For the ease of presentation, we use Latin letters $\mtt{a},\mtt{b},\mtt{c},\dots$ instead of $\mtt{a}_1,\mtt{a}_2,\mtt{a}_3,\dots$ when $k$ is relatively small: $k \le 26$.

\begin{example}
The cryptarithm $\vec{w} = \lrangle{\mtt{send},\mtt{more},\mtt{money}}$ in Example~\ref{ex:sendmoremoney} is not canonical.
Its canonical form is $\vec{v} = \lrangle{\mtt{gbda},\mtt{hfeb},\mtt{hfdbc}}$, whose sequential form is $\psi(\vec{v})=\mtt{abc}\,\mtt{deb}\,\mtt{bfd}\,\mtt{ghf}\,\mtt{\dl\dl h}\,\mtt{\dl\dl\dl}$.
\end{example}

\tabcolsep=1mm
\section{Cryptarithm DFAs}
\subsection{Naive Cryptarithm DFA}
We will define a DFA $M_k$ that accepts all and only canonical cryptarithms that admit solutions.
Our DFA is slightly different from the standard ones.
First, each edge is labeled by a trigram so that letters belonging to the same place will be read at once.
Second, it has two distinguishable accepting states $f_1$ and $f_2$
where cryptarithm sequences with unique and multiple solutions shall be accepted at $f_1$ and at $f_2$, respectively.
Accordingly, our DFA is a sextuple $M_k =  \lrangle{Q, \Sigma_k ,\delta,q_0,f_1,f_2}$ where
 $Q_k$ is the state set, $\delta \colon Q \times (\Sigma_k \cup \{\dl\})^3 \rightharpoonup Q$ is the transition partial function,
 and $f_1$ and $f_2$ are accepting states, which define two languages
\begin{align}
	\notag L_{k,\text{uniq}} &= \{\, w \in (\Sigma_k \cup \{\dl\})^{+} \mid \hat{\delta}(q_0,w) = f_1 \,\}
\\	\label{eq:unique}			&= \{\, \psi(\vec{w}) \in (\Sigma_k \cup \{\dl\})^+ \mid \text{$\vec{w}$ admits exactly one solution}\,\}\,,
\\	\notag	L_{k,\text{multi}} &= \{\, w \in (\Sigma_k \cup \{\dl\})^{+} \mid \hat{\delta}(q_0,w) = f_2 \,\}
\\	\label{eq:multi}			&= \{\, \psi(\vec{w}) \in (\Sigma_k \cup \{\dl\})^+ \mid \text{$\vec{w}$ admits at least two solutions}\,\}\,,
\end{align}
where $\hat{\delta}$ is the usual extension of $\delta$ for domain $((\Sigma_k \cup \{\dl\})^3)^*$.
We call a string $w \in ((\Sigma_k \cup \{\dl\})^3)^*$ 
 \emph{valid} if it is a prefix of some canonical cryptarithm sequence with at least one solution.
We say that an assignment $\theta \colon \rst{\Sigma_k}{w} \to N_k$ is \emph{consistent with $w$}
if there is an extension of $\theta$ which is a solution of a cryptarithm sequence of which $w$ is a prefix.
 When $\rst{\Sigma_k}{w} = \Sigma_{k-1}$, each consistent assignment on $\Sigma_{k-1}$ has just one trivial proper extension injection with domain $\Sigma_k$.
Therefore, we ``promote'' consistent assignments on $\Sigma_{k-1}$ to their extensions on $\Sigma_k$.
We let $\Theta(w)$ denote the set of consistent assignments, possibly with promotion:
\[
	\Theta(w) =\begin{cases}
		\{\, \theta \colon {\Sigma_k} \to N_k \mid \text{$\theta$ is consistent with $w$} \,\}	& \text{if $\rst{\Sigma_k}{w} = \Sigma_{k-1}$,}
\\		\{\, \theta \colon \rst{\Sigma_k}{w} \to N_k \mid \text{$\theta$ is consistent with $w$} \,\}	& \text{otherwise.}
	\end{cases}
\,\]
For a valid sequence $w$, one can characterize succeeding sequences $v$ that will make $wv$ a solvable canonical cryptarithm sequence with $\Theta(w)$ and other parameters.
The parameters the DFA $M_k$ maintains in its states have the form $\lranggle{d_1,d_2, \ell, P}$, which we will call a \emph{configuration}.
Every state except accepting ones has a unique configuration.
Among those parameters, $d_1,d_2 \in \{0,1\}$ are used to ensure that a sequence may be extended to a cryptarithm sequence and
$\ell \in \{1,\dots,k\}$ is used to ensure that a sequence may be extended to a canonical one.
The last parameter $P$ is a non-empty set that remembers possible assignments on letters together with auxiliary information.
Suppose that the configuration of the state $q$ reached from $q_0$ by reading a valid sequence $w$ in $M_k$ is $\lranggle{d_1,d_2, \ell, P}$
and let $w = w' x_1 x_2 x_3$ with $x_1,x_2,x_3 \in \Sigma_k \cup \{\dl\}$ and $\psi^{-1}(w)=\lrangle{w_1,w_2,w_3}$.
Then,
\begin{itemize}
	\item $d_i = 1$ if $x_i = \dl$ and $d_i = 0$ otherwise for $i=1,2$, 
	\item $\ell = \min \{k,\, |\rst{\Sigma}{w}|+1\}$,
	\item $P$ consists of $\lrsquare{\theta,c,b_1,b_2} \in \Theta(w) \times \{0,1\}^3$ where
	\begin{itemize}
		\item $\theta \in \Theta(w)$,
		\item $\hat \theta(w_1) + \hat \theta(w_2) = \hat \theta(w_3) + ck^{|w_3|}$,
		\item $b_i = 0$ if $x_i \neq \dl$ and $\theta(x_i) = 0$, and $b_i = 1$ otherwise, for $i=1,2$.
	\end{itemize}
\end{itemize}
One can see $P$ as a function from $\Theta(w)$ to $\{0,1\}^3$.
For $\lrsquare{\theta,c,b_1,b_2} \in P$, when $c=1$, we have a carry under the assignment $\theta$.
When $b_i=0$, the $i^\text{th}$ term must be extended to have a more significant digit since the current most significant digit is $0$ under $\theta$.

Now let us define  $M_k =  \lrangle{Q,\Sigma_k,\delta,q_0,f_1,f_2}$ so that $M_k$ satisfies the above.
We identify a state and its configuration, since no distinct states have the same configuration: in case two states happen to have the same configuration, they must be merged.
The initial state is the configuration $\lranggle{0,0,1,\{\lrsquare{\varnothing,0,0,0}\}}$, where $\varnothing$ is the empty assignment.

The transition function $\delta$ is defined as follows.
For $x_1,x_2,x_3 \in \Sigma_k \cup \{\dl\}$, let us write $\lranggle{d_1,d_2,\ell,P} \xRightarrow{x_1x_2x_3} \lranggle{d_1',d_2',\ell',P'}$ if
\begin{itemize}
	\item $x_3 = \dl$ implies $x_1=x_2=\dl$,
	\item $d_i=1$ implies $x_i=\dl$ for $i=1,2$,
	\item $d_i' = 1$ if $x_i = \dl$, and $d_i' = 0$ otherwise, for $i=1,2$,
	\item $x_1 \in \Sigma_\ell \cup \{\dl\}$, $x_2 \in \Sigma_{\ell_1} \cup \{\dl\}$, $x_3 \in \Sigma_{\ell_2} \cup \{\dl\}$, where $\ell_1 = \ell$ if $x_1 \in \Sigma_\ell$ and $\ell_1 = \min\{k,\, \ell+1\}$ otherwise, and $\ell_2$ is defined from $\ell_1$ and $x_2$ in the same manner,
	\item $\ell'$ is defined from $\ell_2$ and $x_3$ in the same manner,
	\item $P' = \{\, p'  \mid p \xrightarrow{x_1x_2x_3} p' \text{ for some } p \in P\,\}$ is not empty,
\end{itemize}
 where we write $\lrsquare{\theta,c,b_1,b_2} \xrightarrow{x_1x_2x_3} \lrsquare{\theta',c',b_1',b_2'}$ if
\begin{itemize}
	\item $b_i=0$ implies $x_i \neq \dl$ for $i=1,2$,
	\item $\theta'\colon \Sigma' \to N_k$ is an extension of $\theta$ where $\Sigma'=\Sigma_k$ if $\ell'=k$, and $\Sigma' = \Sigma_{\ell'-1}$ otherwise,
	\item $c+\tilde{\theta}'(x_1)+\tilde{\theta}'(x_2) = c'k + \tilde{\theta}'(x_3)$ where $\tilde{\theta}'$ extends $\theta'$ by $\tilde{\theta}'(\dl)=0$,
	\item $b_i' = 0$ if $x_i \neq \dl$ and $\theta'(x_i)=0$, and $b_i'=1$ otherwise, for $i=1,2$.
\end{itemize}
If $x_1x_2x_3 \neq \dl\dl\dl$, then we define $\delta(q,x_1x_2x_3) = q'$ for $q \xRightarrow{x_1x_2x_3} q'$. 
When $x_1x_2x_3 = \dl\dl\dl$, this means the end of the input sequence, if it is a cryptarithm sequence.
For $q'=\lrangle{d_1',d_2',\ell',P'}$ with $q \xRightarrow{\dl\dl\dl} q'$, we define
$\delta(q,\dl\dl\dl) = f_1$ if $|P'| = 1$,
and
$\delta(q,\dl\dl\dl) = f_2$ if $|P'| \ge 2$.

The state set $Q$ is defined to consist of the states reachable from the initial state according to $\delta$.

\begin{example}
Let $k=3$.
Suppose that a state $q$ in $M_3$ has a configuration $\lranggle{d_1,d_2,\ell,P}=\lranggle{0,0,2,P}$ with
\[
	P = \{\, \lrsquare{\{\mtt{a} \mapsto 0\},c,b_1,b_2} \,\}= \{\, \lrsquare{\{\mtt{a} \mapsto 0\},0,0,0} \,\}
\,.\]
In fact, this state is reached by reading $\mtt{aaa}$ from the initial state,
where we did not yet find $\dl$ (so $d_1=d_2=0$), the second letter $\mtt{b}$ may appear in the nearest future (so $\ell=2$), and the only consistent assignment $\theta$ maps $\mtt{a}$ to $0$ (otherwise $\theta(\mtt{a})+\theta(\mtt{a}) \neq \theta(\mtt{a})$), under which we have no carry ($c=0$), but each term must not finish ($b_1=b_2=0$).
Therefore, this state $q$ has no outgoing transition edge labeled with a trigram including $\dl$.
When reading $\mtt{aaa}$ again from this state, the situation does not change.
So we have $\delta(q,\mtt{aaa})=q$.
If we read $\mtt{abb}$, where $\mtt{b}$ is a new letter, we reach a new state $q'$.
Although the last letter $\mtt{c}$ in $\Sigma_3$ has not appeared yet, it is ready to come.
The domain of the assignments in the configuration of $q'$ is now $\Sigma_3$.
We have two consistent assignments extending the one $\{\mtt{a} \mapsto 0\}$ in $q$.
One maps $\mtt{b}$ to $1$ and the other maps $\mtt{b}$ to $2$.
In both cases, we have no carry and the second term may finish.
Thus, the configuration of $q'$ is $\lranggle{0,0,3,P'}$ with
\[
	P' = \{\, \lrsquare{\{\mtt{a} \mapsto 0,\,\mtt{b} \mapsto 1,\,\mtt{c} \mapsto 2\}, 0,0,1},\ \lrsquare{\{\mtt{a} \mapsto 0,\,\mtt{b} \mapsto 2,\,\mtt{c} \mapsto 1\}, 0,0,1} \,\}
\,.\]
On the other hand, it is not hard to see that there is no $p''$ such that $\lrsquare{\{\mtt{a} \mapsto 0\}, 0,0,0} \xrightarrow{\mtt{abc}} p''$.
Hence $q$ has no edge labeled with $\mtt{abc}$.
In this way, we decide whether a state has an outgoing edge labeled with a trigram over $\Sigma_k \cup \{\dl\}$ and the configuration of the reached state.
\end{example}

We have now established Equations~(\ref{eq:unique}) and~(\ref{eq:multi}).
An assignment $\theta$ is a solution of a cryptarithm sequence $w\dl\dl\dl$ if and only if
$\lrsquare{\theta,0,1,1} \in P$ of the configuration $\lranggle{d_1,d_2,\ell,P}$ of the state $\delta(q_0,w)$.
In other words, one can regard our DFA as a Mealy machine that outputs solutions when reading $\dl\dl\dl$.

We remark that the constructed DFA is minimum as a Mealy machine but is not necessarily minimum if we ignore output solutions.
For example, let us consider the states reached by $\mtt{abc \dl ab}$ and $\mtt{abc \dl ba}$ from the initial state in $M_3$.
They have different configurations $\lranggle{1,0,3,P_1}$ and $\lranggle{1,0,3,P_2}$ where
\(
	P_i = \{\, \lrsquare{ \{\mtt{a} \mapsto i,\,\mtt{b} \mapsto (3-i),\,\mtt{c} \mapsto 0 \},0,1,1 } \,\}
\text{ for $i=1,2$.}
\)
Those states are not merged but the strings that will lead us to the accepting state $f_1$ from those states coincide; namely, they have the form ${\dl x_1x_1} \dots {\dl x_nx_n}\dl\dl\dl$ where $x_i \in \{\mtt{a},\mtt{b},\mtt{c}\}$ for $i < n$, $x_n \in \{\mtt{a},\mtt{b}\}$ and $n \ge 0$.

The number of states of $M_k$ is bounded by the number of possible configurations.
A trivial and loose upper bound on it is $2^{\mrm{O}(k!)}$.
If one is interested only in cryptarithms with a unique solution, one can remove the state $f_2$.
If uniqueness of a solution does not matter, two accepting states $f_1$ and $f_2$ can be merged.

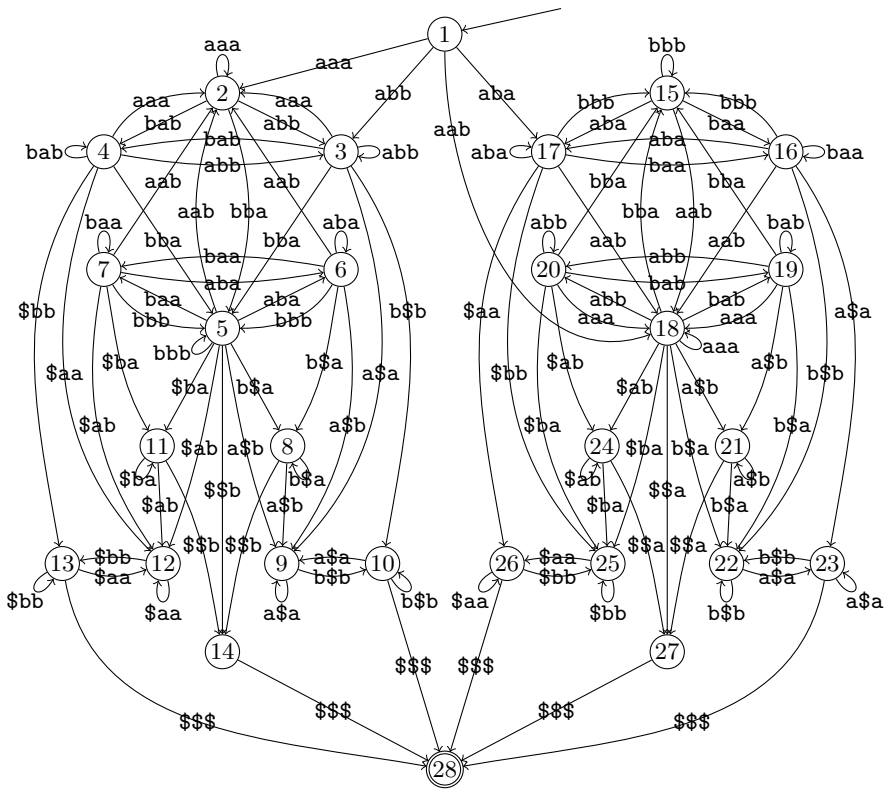
\begin{figure}
	\centering
		  \begin{tikzpicture}[scale=0.78, every node/.style={circle},inner sep=0pt,minimum size=4.5mm]\small
		  	\node[] (0) at (8.0,17.5) {}; 
		  	\node[draw] (1) at (5.75,17.0) {$1$}; 
		  	
		  	\node[draw] (2) at (2,16) {$2$}; 
		    \node[draw] (3) at (4,15) {$3$}; 
		    \node[draw] (4) at (0,15) {$4$}; 
		    \node[draw] (5) at (2,12) {$5$};
		    \node[draw] (6) at (3.0,8) {$9$}; 
		    \node[draw] (7) at (4.7,8) {$10$};
		    \node[draw] (8) at (1.0,8) {$12$}; 
		    \node[draw] (9) at (-0.7,8) {$13$}; 
		    \node[draw] (10) at (4,13) {$6$};
		    \node[draw] (11) at (0,13) {$7$}; 
		    \node[draw] (12) at (3.1,10) {$8$}; 
		    \node[draw] (13) at (0.9,10) {$11$}; 
		    \node[draw] (14) at (2,6.5) {$14$}; 
		    
		    \node[draw] (15) at (9.5,16) {$15$};
		    \node[draw] (16) at (11.5,15) {$16$};
		    \node[draw] (17) at (7.5,15) {$17$};
		    \node[draw] (18) at (9.5,12) {$18$};
		    \node[draw] (19) at (10.5,8) {$22$};
		    \node[draw] (20) at (12.2,8) {$23$};
		    \node[draw] (21) at (8.5,8) {$25$};
		    \node[draw] (22) at (6.8,8) {$26$};
		    \node[draw] (23) at (11.5,13) {$19$};
		    \node[draw] (24) at (7.5,13) {$20$};
		    \node[draw] (25) at (10.6,10) {$21$};
		    \node[draw] (26) at (8.4,10) {$24$};
		    \node[draw] (27) at (9.5,6.5) {$27$};
		    
		    \node[accepting, draw] (28) at (5.75,4.5) {$28$};
			\draw[->] (0) to (1);
		    \draw[->] (27) to node[] {$\mtt{\dl\dl{}\dl{}}$} (28);
		    
		    \draw[->] (26) to node[] {$\mtt{\dl{}ba}$} (21);
		    \draw[->,out=225,in=255,loop] (26) to node[] {$\mtt{\dl{}ab}$} (26);
		    \draw[->,out=300,in=100] (26) to node[] {$\mtt{\dl\dl{}a}$} (27);
		    
		    \draw[->] (25) to node[] {$\mtt{b\dl{}a}$} (19);
		    \draw[->,out=315,in=285,loop] (25) to node[] {$\mtt{a\dl{}b}$} (25);
		    \draw[->,out=240,in=80] (25) to node[] {$\mtt{\dl\dl{}a}$} (27);
		    
		    \draw[->] (24) to node[] {$\mtt{bba}$} (15);
		    \draw[->,out=350,in=190] (24) to node[] {$\mtt{bab}$} (23);
		    \draw[->, loop above] (24) to node[above=-4pt] {$\mtt{abb}$} (24);
		    \draw[->,out=300,in=180] (24) to node[] {$\mtt{aaa}$} (18);
		    \draw[->,out=260,in=125] (24) to node[] {$\mtt{\dl{}ba}$} (21);
		    \draw[->,out=280,in=120] (24) to node[] {$\mtt{\dl{}ab}$} (26);
		    
		    \draw[->] (23) to node[] {$\mtt{bba}$} (15);
		    \draw[->, loop above] (23) to node[above=-4pt] {$\mtt{bab}$} (23);
		    \draw[->,out=280,in=55] (23) to node[] {$\mtt{b\dl{}a}$} (19);
		    \draw[->,out=170,in=10] (23) to node[] {$\mtt{abb}$} (24);
		    \draw[->,out=240,in=0] (23) to node[] {$\mtt{aaa}$} (18);
		    \draw[->,out=260,in=60] (23) to node[] {$\mtt{a\dl{}b}$} (25);
		    
		    \draw[->,out=340,in=200] (22) to node[] {$\mtt{\dl{}bb}$} (21);
		    \draw[->,out=210,in=240,loop] (22) to node[below left=-2pt] {$\mtt{\dl{}aa}$} (22);
		    \draw[->] (22) to node[] {$\mtt{\dl\dl{}\dl{}}$} (28);
		    
		    \draw[->, loop below] (21) to node[below=-4pt] {$\mtt{\dl{}bb}$} (21);
		    \draw[->,out=170,in=10] (21) to node[] {$\mtt{\dl{}aa}$} (22);
		    
		    \draw[->,out=170,in=10] (20) to node[] {$\mtt{b\dl{}b}$} (19);
		    \draw[->,out=300,in=330, loop] (20) to node[below right=-2pt] {$\mtt{a\dl{}a}$} (20);
		    \draw[->,out=260,in=10] (20) to node[] {$\mtt{\dl\dl{}\dl{}}$} (28);
		    
		    \draw[->, loop below] (19) to node[below=-4pt] {$\mtt{b\dl{}b}$} (19);
		    \draw[->,out=340,in=200] (19) to node[] {$\mtt{a\dl{}a}$} (20);
		    
		    \draw[->,out=110,in=250] (18) to node[] {$\mtt{bba}$} (15);
		    \draw[->] (18) to node[] {$\mtt{bab}$} (23);
		    \draw[->,out=280,in=110] (18) to node[] {$\mtt{b\dl{}a}$} (19);
		    \draw[->] (18) to node[] {$\mtt{abb}$} (24);
		    \draw[->,out=315,in=345, loop] (18) to node[right] {$\mtt{aaa}$} (18);
		    \draw[->] (18) to node[] {$\mtt{a\dl{}b}$} (25);
		    \draw[->,out=260,in=70] (18) to node[] {$\mtt{\dl{}ba}$} (21);
		    \draw[->] (18) to node[] {$\mtt{\dl{}ab}$} (26);
		    \draw[->] (18) to node[] {$\mtt{\dl\dl{}a}$} (27);
		    
		    \draw[->,out=60,in=180] (17) to node[] {$\mtt{bbb}$} (15);
		    \draw[->,out=350,in=190] (17) to node[] {$\mtt{baa}$} (16);
		    \draw[->, loop left] (17) to node[] {$\mtt{aba}$} (17);
		    \draw[->] (17) to node[] {$\mtt{aab}$} (18);
		    \draw[->,out=250,in=135] (17) to node[] {$\mtt{\dl{}bb}$} (21); 
		    \draw[->,out=235,in=100] (17) to node[pos=0.4] {$\mtt{\dl{}aa}$} (22);
		    
		    \draw[->,out=120,in=0] (16) to node[] {$\mtt{bbb}$} (15);
		    \draw[->, loop right] (16) to node[] {$\mtt{baa}$} (16);
		    \draw[->,out=170,in=10] (16) to node[] {$\mtt{aba}$} (17);
		    \draw[->] (16) to node[] {$\mtt{aab}$} (18);
		    \draw[->,out=290,in=45] (16) to node[] {$\mtt{b\dl{}b}$} (19);
		    \draw[->,out=305,in=80] (16) to node[pos=0.4] {$\mtt{a\dl{}a}$} (20);
		    		    
		    \draw[->, loop above] (15) to node[above=-4pt] {$\mtt{bbb}$} (15);
		    \draw[->] (15) to node[] {$\mtt{baa}$} (16);
		    \draw[->] (15) to node[] {$\mtt{aba}$} (17);
		    \draw[->,out=290,in=70] (15) to node[] {$\mtt{aab}$} (18);

			\draw[->,out=270,in=200] (1) to node[pos=0.18] {$\mtt{aab}$} (18);
			\draw[->] (1) to node[] {$\mtt{aba}$} (17);

		    \draw[->] (14) to node[] {$\mtt{\dl\dl{}\dl{}}$} (28);
		    
		    \draw[->] (13) to node[] {$\mtt{\dl{}ab}$} (8);
		    \draw[->,out=225,in=255,loop] (13) to node[] {$\mtt{\dl{}ba}$} (13);
		    \draw[->,out=300,in=100] (13) to node[] {$\mtt{\dl\dl{}b}$} (14);
		    
		    \draw[->] (12) to node[] {$\mtt{a\dl{}b}$} (6);
		    \draw[->,out=315,in=285,loop] (12) to node[] {$\mtt{b\dl{}a}$} (12);
		    \draw[->,out=240,in=80] (12) to node[] {$\mtt{\dl\dl{}b}$} (14);
		    
		    \draw[->] (11) to node[] {$\mtt{aab}$} (2);
		    \draw[->,out=350,in=190] (11) to node[] {$\mtt{aba}$} (10);
		    \draw[->, loop above] (11) to node[above=-4pt] {$\mtt{baa}$} (11);
		    \draw[->,out=300,in=180] (11) to node[] {$\mtt{bbb}$} (5);
		    \draw[->,out=260,in=125] (11) to node[] {$\mtt{\dl{}ab}$} (8);
		    \draw[->,out=280,in=120] (11) to node[] {$\mtt{\dl{}ba}$} (13);
		    
		    \draw[->] (10) to node[] {$\mtt{aab}$} (2);
		    \draw[->, loop above] (10) to node[above=-4pt] {$\mtt{aba}$} (10);
		    \draw[->,out=280,in=55] (10) to node[] {$\mtt{a\dl{}b}$} (6);
		    \draw[->,out=170,in=10] (10) to node[] {$\mtt{baa}$} (11);
		    \draw[->,out=240,in=0] (10) to node[] {$\mtt{bbb}$} (5);
		    \draw[->,out=260,in=60] (10) to node[] {$\mtt{b\dl{}a}$} (12);
		    
		    \draw[->,out=340,in=200] (9) to node[] {$\mtt{\dl{}aa}$} (8);
		    \draw[->,out=210,in=240, loop] (9) to node[below left=-2pt] {$\mtt{\dl{}bb}$} (9);
		    \draw[->,out=280,in=170] (9) to node[] {$\mtt{\dl\dl{}\dl{}}$} (28);
		    
		    \draw[->, loop below] (8) to node[below=-4pt] {$\mtt{\dl{}aa}$} (8);
		    \draw[->,out=170,in=10] (8) to node[] {$\mtt{\dl{}bb}$} (9);
		    
		    \draw[->,out=170,in=10] (7) to node[] {$\mtt{a\dl{}a}$} (6);
		    \draw[->,out=300,in=330, loop] (7) to node[below right=-2pt] {$\mtt{b\dl{}b}$} (7);
		    \draw[->] (7) to node[] {$\mtt{\dl\dl{}\dl{}}$} (28);
		    
		    \draw[->, loop below] (6) to node[below=-4pt] {$\mtt{a\dl{}a}$} (6);
		    \draw[->,out=340,in=200] (6) to node[] {$\mtt{b\dl{}b}$} (7);
		    
		    \draw[->,out=110,in=250] (5) to node[] {$\mtt{aab}$} (2);
		    \draw[->] (5) to node[] {$\mtt{aba}$} (10);
		    \draw[->,out=280,in=110] (5) to node[] {$\mtt{a\dl{}b}$} (6);
		    \draw[->] (5) to node[] {$\mtt{baa}$} (11);
		    \draw[->,out=235,in=205, loop] (5) to node[left] {$\mtt{bbb}$} (5);
		    \draw[->] (5) to node[] {$\mtt{b\dl{}a}$} (12);
		    \draw[->,out=260,in=70] (5) to node[] {$\mtt{\dl{}ab}$} (8);
		    \draw[->] (5) to node[] {$\mtt{\dl{}ba}$} (13);
		    \draw[->] (5) to node[] {$\mtt{\dl\dl{}b}$} (14);
		    
		    \draw[->,out=60,in=180] (4) to node[] {$\mtt{aaa}$} (2);
		    \draw[->,out=350,in=190] (4) to node[] {$\mtt{abb}$} (3);
		    \draw[->, loop left] (4) to node[] {$\mtt{bab}$} (4);
		    \draw[->] (4) to node[] {$\mtt{bba}$} (5);
		    \draw[->,out=250,in=135] (4) to node[] {$\mtt{\dl{}aa}$} (8); 
		    \draw[->,out=235,in=100] (4) to node[pos=0.4] {$\mtt{\dl{}bb}$} (9);
		    
		    \draw[->,out=120,in=0] (3) to node[] {$\mtt{aaa}$} (2);
		    \draw[->, loop right] (3) to node[] {$\mtt{abb}$} (3);
		    \draw[->,out=170,in=10] (3) to node[] {$\mtt{bab}$} (4);
		    \draw[->] (3) to node[] {$\mtt{bba}$} (5);
		    \draw[->,out=290,in=45] (3) to node[] {$\mtt{a\dl{}a}$} (6);
		    \draw[->,out=305,in=80] (3) to node[pos=0.4] {$\mtt{b\dl{}b}$} (7);
		    		    
		    \draw[->, loop above] (2) to node[above=-4pt] {$\mtt{aaa}$} (2);
		    \draw[->] (2) to node[] {$\mtt{abb}$} (3);
		    \draw[->] (2) to node[] {$\mtt{bab}$} (4);
		    \draw[->,out=290,in=70] (2) to node[] {$\mtt{bba}$} (5);

		    \draw[->] (1) to node[] {$\mtt{aaa}$} (2);
		    \draw[->] (1) to node[] {$\mtt{abb}$} (3);
	
		  \end{tikzpicture}
	\caption{DFA $M_2$ that accepts binary solvable canonical cryptarithm sequences. The initial state is $q_0=1$ and the accepting state is $f_1=28$.  The other accepting state $f_2$ is missing in $M_2$.}
	\label{fig:dfa_base2}
\end{figure}

Figure~\ref{fig:dfa_base2} shows the finally obtained automaton for $k=2$.
This automaton $M_2$ misses the accepting state $f_2$, because no cryptarithm has two distinct binary solutions.

\begin{figure}
	\centering
	\newcommand{\pv}[1]{\linethickness{0.07mm}\framebox(15,7.5){\color{red}$\mathtt{#1}$}}
		  \begin{tikzpicture}[scale=0.78, every node/.style={circle},inner sep=0pt,minimum size=4.5mm]\small
		  	\node[] (0) at (8.0,17.5) {}; 
		  	\node[draw] (1) at (5.75,17.0) {$1$}; 
		  	
		  	\node[draw] (2) at (2,16) {$2$}; 
		    \node[draw] (3) at (4,15) {$3$}; 
		    \node[draw] (4) at (0,15) {$4$}; 
		    \node[draw] (5) at (2,12) {$5$};
		    \node[draw] (6) at (3.0,8) {$9$}; 
		    \node[draw] (7) at (4.7,8) {$10$};
		    \node[draw] (8) at (1.0,8) {$12$}; 
		    \node[draw] (9) at (-0.7,8) {$13$}; 
		    \node[draw] (10) at (4,13) {$6$};
		    \node[draw] (11) at (0,13) {$7$}; 
		    \node[draw] (12) at (3.1,10) {$8$}; 
		    \node[draw] (13) at (0.9,10) {$11$}; 
		    \node[draw] (14) at (2,6.5) {$14$}; 
		    		    
		    \node[accepting, draw] (28) at (5.75,4.5) {$28$};
			\draw[->] (0) to (1);

			\draw[->,out=270,in=340,red] (1) to node[pos=0.16,right] {\pv{aab}} (5);
			\draw[->,out=180,in=90,red] (1) to node[pos=0.25,above=-4pt] {\pv{aba}} (4);

		    \draw[->] (14) to node[] {$\mtt{\dl\dl{}\dl{}}$} (28);
		    
		    \draw[->] (13) to node[] {$\mtt{\dl{}ab}$} (8);
		    \draw[->,out=225,in=255,loop] (13) to node[] {$\mtt{\dl{}ba}$} (13);
		    \draw[->,out=300,in=100] (13) to node[] {$\mtt{\dl\dl{}b}$} (14);
		    
		    \draw[->] (12) to node[] {$\mtt{a\dl{}b}$} (6);
		    \draw[->,out=315,in=285,loop] (12) to node[] {$\mtt{b\dl{}a}$} (12);
		    \draw[->,out=240,in=80] (12) to node[] {$\mtt{\dl\dl{}b}$} (14);
		    
		    \draw[->] (11) to node[] {$\mtt{aab}$} (2);
		    \draw[->,out=350,in=190] (11) to node[] {$\mtt{aba}$} (10);
		    \draw[->, loop above] (11) to node[above=-4pt] {$\mtt{baa}$} (11);
		    \draw[->,out=300,in=180] (11) to node[] {$\mtt{bbb}$} (5);
		    \draw[->,out=260,in=125] (11) to node[] {$\mtt{\dl{}ab}$} (8);
		    \draw[->,out=280,in=120] (11) to node[] {$\mtt{\dl{}ba}$} (13);
		    
		    \draw[->] (10) to node[] {$\mtt{aab}$} (2);
		    \draw[->, loop above] (10) to node[above=-4pt] {$\mtt{aba}$} (10);
		    \draw[->,out=280,in=55] (10) to node[] {$\mtt{a\dl{}b}$} (6);
		    \draw[->,out=170,in=10] (10) to node[] {$\mtt{baa}$} (11);
		    \draw[->,out=240,in=0] (10) to node[] {$\mtt{bbb}$} (5);
		    \draw[->,out=260,in=60] (10) to node[] {$\mtt{b\dl{}a}$} (12);
		    
		    \draw[->,out=340,in=200] (9) to node[] {$\mtt{\dl{}aa}$} (8);
		    \draw[->,out=210,in=240, loop] (9) to node[below left=-2pt] {$\mtt{\dl{}bb}$} (9);
		    \draw[->,out=280,in=170] (9) to node[] {$\mtt{\dl\dl{}\dl{}}$} (28);
		    
		    \draw[->, loop below] (8) to node[below=-4pt] {$\mtt{\dl{}aa}$} (8);
		    \draw[->,out=170,in=10] (8) to node[] {$\mtt{\dl{}bb}$} (9);
		    
		    \draw[->,out=170,in=10] (7) to node[] {$\mtt{a\dl{}a}$} (6);
		    \draw[->,out=300,in=330, loop] (7) to node[below right=-2pt] {$\mtt{b\dl{}b}$} (7);
		    \draw[->] (7) to node[] {$\mtt{\dl\dl{}\dl{}}$} (28);
		    
		    \draw[->, loop below] (6) to node[below=-4pt] {$\mtt{a\dl{}a}$} (6);
		    \draw[->,out=340,in=200] (6) to node[] {$\mtt{b\dl{}b}$} (7);
		    
		    \draw[->,out=110,in=250] (5) to node[] {$\mtt{aab}$} (2);
		    \draw[->] (5) to node[] {$\mtt{aba}$} (10);
		    \draw[->,out=280,in=110] (5) to node[] {$\mtt{a\dl{}b}$} (6);
		    \draw[->] (5) to node[] {$\mtt{baa}$} (11);
		    \draw[->,out=235,in=205, loop] (5) to node[left] {$\mtt{bbb}$} (5);
		    \draw[->] (5) to node[] {$\mtt{b\dl{}a}$} (12);
		    \draw[->,out=260,in=70] (5) to node[] {$\mtt{\dl{}ab}$} (8);
		    \draw[->] (5) to node[] {$\mtt{\dl{}ba}$} (13);
		    \draw[->] (5) to node[] {$\mtt{\dl\dl{}b}$} (14);
		    
		    \draw[->,out=60,in=180] (4) to node[] {$\mtt{aaa}$} (2);
		    \draw[->,out=350,in=190] (4) to node[] {$\mtt{abb}$} (3);
		    \draw[->, loop left] (4) to node[] {$\mtt{bab}$} (4);
		    \draw[->] (4) to node[] {$\mtt{bba}$} (5);
		    \draw[->,out=250,in=135] (4) to node[] {$\mtt{\dl{}aa}$} (8); 
		    \draw[->,out=235,in=100] (4) to node[pos=0.4] {$\mtt{\dl{}bb}$} (9);
		    
		    \draw[->,out=120,in=0] (3) to node[] {$\mtt{aaa}$} (2);
		    \draw[->, loop right] (3) to node[] {$\mtt{abb}$} (3);
		    \draw[->,out=170,in=10] (3) to node[] {$\mtt{bab}$} (4);
		    \draw[->] (3) to node[] {$\mtt{bba}$} (5);
		    \draw[->,out=290,in=45] (3) to node[] {$\mtt{a\dl{}a}$} (6);
		    \draw[->,out=305,in=80] (3) to node[pos=0.4] {$\mtt{b\dl{}b}$} (7);
		    		    
		    \draw[->, loop above] (2) to node[above=-4pt] {$\mtt{aaa}$} (2);
		    \draw[->] (2) to node[] {$\mtt{abb}$} (3);
		    \draw[->] (2) to node[] {$\mtt{bab}$} (4);
		    \draw[->,out=290,in=70] (2) to node[] {$\mtt{bba}$} (5);

		    \draw[->] (1) to node[] {$\mtt{aaa}$} (2);
		    \draw[->] (1) to node[] {$\mtt{abb}$} (3);
	
		  \end{tikzpicture}
	\caption{DFA with permutation edges for binary solvable canonical cryptarithm sequences,
	 where black trigram labels are with the identity permutation $\iota$ and red boxed trigram labels are with $\{\mtt{a} \mapsto \mtt{b}, \mtt{b} \mapsto \mtt{a}\}$.}
	\label{fig:dfa_compressed}
\end{figure}%


\subsection{Compressed Cryptarithm DFA}
By observing Fig.~\ref{fig:dfa_base2}, one may realize that the DFA has isomorphic substructures.
Namely, the sub-automaton $M_2^{2}$ whose initial state is set to be $2$ is isomorphic to $M_2^{15}$ with initial state $15$
by swapping $\mtt{a}$ and $\mtt{b}$ on the edge labels.
There exist just two binary assignments, $\{\mtt{a} \mapsto 0,\mtt{b} \mapsto 1\}$ and $\{\mtt{a} \mapsto 1,\mtt{b} \mapsto 0\}$.
The first trigram of any cryptarithm sequence uniquely determines one of the two as a consistent assignment.
The former assignment corresponds to $M_2^{2}$ and the latter to $M_2^{15}$.
We say that two configurations $\lranggle{ d_1,d_2,\ell,P }$ and  $\lranggle{ d_1',d_2',\ell',P' }$ are \emph{permutative variants}
if  $d_1=d_1'$, $d_2=d_2'$, $\ell=\ell'$, and there is a bijection $\pi$ on $\Sigma_{m}$ with $m = k$ if $\ell = k$ and $m=\ell-1$ otherwise such that
\[
	P' = \pi(P) = \{\, \lrsquare{ \theta \circ \pi, c, b_1, b_2 } \mid \lrsquare{ \theta, c, b_1, b_2 } \in P\,\}
\,.\]
Clearly if the configurations of two states are permutative variants, the subautomata consisting of reachable states from those states are isomorphic under the permutation.
This allows us to reduce the size of the automaton by merging those states.
In our new DFA $\widetilde{M}_k$, each transition edge has two labels: one is a trigram as before and the other is a permutation on $\Sigma_k$.
After passing a transition edge labeled with a permutation $\pi$, we will follow transition edges by replacing each letter in accordance with $\pi$.
Figure~\ref{fig:dfa_compressed} shows $\widetilde{M}_2$.

We formally define this new kind of DFAs with edges labeled with a letter and a permutation.
A \emph{DFA with permutation edges} is a sextuple $M = \lrangle{Q,\Sigma,\delta,\gamma,q_0,F}$, where $\delta$ and $\gamma$ are partial functions $Q \times \Sigma \rightharpoonup Q$ and $Q \times \Sigma \rightharpoonup \Pi_\Sigma$, respectively, where $\Pi_\Sigma$ is the set of all permutations over $\Sigma$, such that the domains of $\delta$ and $\gamma$ coincide.
For $x \in \Sigma$, $w \in \Sigma^*$ and $q \in Q$, define $\hat{\delta}\colon Q \times \Sigma^* \rightharpoonup Q$ and $\hat{\gamma}\colon Q \times \Sigma^* \rightharpoonup \Pi_\Sigma$ by
\begin{align*}
	\hat{\delta}(q,\varepsilon) &= q,
\\	\hat{\gamma}(q,\varepsilon) &= \iota,
\\	\hat{\delta}(q,wx) &= \delta(\hat{\delta}(q,w), \hat{\gamma}(q,w)(x)),
\\	\hat{\gamma}(q,wx) &= \gamma(\hat{\delta}(q,w), \hat{\gamma}(q,w)(x)) \circ \hat{\gamma}(q,w),
\end{align*}
where $\iota$ is the identity.
The strings that the automaton $M$ accepts are those $w \in \Sigma^*$ such that $\hat\delta(q_0,w) \in F$.
Mealy machines with permutation edges can also be defined, where outputs may depend on the current state, permutation and next input letter.

We modify $M_k$ to $\widetilde{M}_k$ by merging states that are permutative variants into a representative and adding appropriate permutation labels to edges.
We choose as representative the state that is lexicographically earliest among permutative variants with respect to their representations in the implementation.
In our cryptarithm DFAs with permutation edges, permutation labels are defined on letters in $\Sigma_k$ and homomorphically extended to trigrams on $\Sigma_k \cup \{\dl\}$, where $\dl$ is always mapped to $\dl$ itself.
Algorithm~\ref{alg:dfaconstruction} shows the pseudo code for constructing $\widetilde{M}_k$.
An assignment $\theta\circ \hat\gamma(q_0,w)$ is a solution of a cryptarithm sequence $w\dl\dl\dl$ if and only if
$\lrsquare{\theta ,0,1,1} \in P$ for the configuration $\lranggle{d_1,d_2,\ell,P}$ of the state $\hat\delta(q_0,w)$.

\begin{algorithm}[h]
\caption{Constructing $\widetilde{M}_k$}
\label{alg:dfaconstruction}
	\begin{algorithmic}[1]
	\STATE{let $q_0 := \lranggle{0,0,1,\{\lrsquare{\varnothing,0,0,0}\}}$ and $Q := \{q_0,f_1,f_2\}$;}
	\STATE{push $q_0$ to the stack;}
	\WHILE{the stack is not empty}
	\STATE{pop the top element $q$ from the stack;}
		\FOR{each trigram $u$ on $\Sigma_k \cup \{\dl\}$}
			\IF{there is a configuration $q'$ such that $q \xRightarrow{u} q'$}
				\IF{$u=\dl\dl\dl$}
					\IF{$q'=\lranggle{d_1,d_2,\ell,P}$ with $|P| = 1$}
						\STATE{add an edge from $q$ to $f_1$ with label $\lrangle{u,\iota}$;}
					\ELSE
						\STATE{add an edge from $q$ to $f_2$ with label $\lrangle{u,\iota}$;}
					\ENDIF
				\ELSE
					\STATE{let $\pi \in \Pi_{\Sigma_k}$ be such that $\pi(q') = q''$ is the canonical form of $q'$;}
					\IF{$q'' \notin Q$}
						\STATE{add $q''$ to $Q$ and push $q''$ to the stack;}
					\ENDIF
					\STATE{add an edge from $q$ to $q''$ with label $\lrangle{u,\pi}$;}
				\ENDIF
			\ENDIF
		\ENDFOR
	\ENDWHILE
    \RETURN $\lrangle{Q,\Sigma_k,\delta,\gamma,q_0,f_1,f_2}$;
	\end{algorithmic}
\end{algorithm}

\subsection{Comparison of Naive and Compressed Cryptarithm DFAs}
Table~\ref{tab:fukumenDFA} compares the numbers of states and edges of $M_k$ and $\widetilde{M}_k$.
We succeeded in calculating the automata for $k \le 7$ but gave up for $k \ge 8$ due to the long time calculation and big memory consumption.
For the purpose of reference, we also show the number of states of $\min(M_k)$, the minimized version of $M_k$.
Note that minimization loses the information of possible solutions for cryptarithm sequences and therefore $\min(M_k)$ cannot be used as a solver.
Our compression technique achieves a more compact representation than the classical state minimization technique for solvable cryptarithm sequences, while keeping the solver function.
Table~\ref{tab:fukumenDFA_time} compares the time and space used to construct $M_k$ and $\widetilde{M}_k$.
Our implementation was compiled with Go 1.10 on Ubuntu 14.04 LTS with CPU Xeon E5-2609 2.4GHz and 256 GB memory.
To construct $\widetilde{M}_k$ requires much shorter time and smaller memory than $M_k$ for all $k$.
\begin{table}[H]
	\centering
	\caption{Numbers of states and edges of cryptarithm automata\label{tab:fukumenDFA}}
	\begin{tabular}{c|c|rrrrrrr}
		\multicolumn{1}{c}{} & Base $k$ & 2 & 3 & 4 & 5 & 6 & 7\\ \hline
		&$M_k$ & 28 & 110 & 859 & 10267 & 370719 & 30909627 \\
	States	&$\widetilde{M}_k$ & 15 & 27 & 163 & 1061 & 17805 & 472518 \\
		&$\min(M_k)$ & 27 & 93 & 607 & 6589 & 248192 & -- \\ \hline
		&$M_k$ & 112 & 1032 & 17662 & 350019 & 23508141 & 3017993409 \\
	Edges	&$\widetilde{M}_k$ & 58 & 233 & 3860 & 40042 & 1214972 & 48635469 \\
		&$\min(M_k)$ & 111 & 985 & 16602 & 330297 & 22673144 & -- 
	\end{tabular}
\end{table}

\begin{table}[H]
	\centering
	\caption{Used computational resources for constructing cryptarithm automata}
	\begin{tabular}{c|c|rrrrrr}
	\multicolumn{1}{c}{} &	Base $k$ 			&	2	&	3		&	4	& 5	 & 6 & 7  \\ \hline
	Time &	$M_k$				& $<0.01$ & $<0.01$ & 0.03 & 0.59 & 35 & 87\,min. \\
	(sec.) &	$\widetilde{M}_k$	& $<0.01$ & $<0.01$ & $<0.01$ & $0.15$ & $4.0$ & $149$\,sec. \\ \hline
	Space &	$M_k$				& $<2$ & $<2$ & $4.6$ & $20$ & $1.0$\,GB & $91$\,GB \\ 
	(MB) &	$\widetilde{M}_k$	& $<2$ & $<2$ & $3.1$ & $6.6$ & $80$\,MB & $3.1$\,GB \\
	\end{tabular}
	\label{tab:fukumenDFA_time}
\end{table}

\newpage

\subsection{Cryptarithms with Limited Number of Letters}
As we have observed in the previous subsection, we were unable to compute $M_k$ and $\widetilde{M}_k$ for $k \ge 8$.
On the other hand, there are many interesting decimal cryptarithms in the real world that do not involve all the ten numerals.
It is still interesting to construct a DFA $\widetilde{M}_{k,s}$ that accepts all and only base-$k$ solvable cryptarithm sequences over $\Sigma_s$ for $s \le k$.
This can be achieved by a slight modification on Algorithm~\ref{alg:dfaconstruction},
 where we refrain from making transition edges whose label includes forbidden letters not in $\Sigma_s$.
In addition, when $s=k-1$, we need to give up ``promotion'' of an assignment with domain $\Sigma_{k-1}$ to its extension with domain $\Sigma_k$.
This results actually in a simpler construction algorithm.

Tables~\ref{tab:fukumenDFA_ks} and~\ref{tab:fukumenDFA_ks_time} show the numbers of states and the computation times of the construction of $\widetilde{M}_{k,s}$ for $7 \le k \le 10$ and $2 \le s \le 6$.

\begin{table}
\begin{minipage}[h]{.43\textwidth}
	\centering
  \caption{Numbers of states of $\widetilde{M}_{k,s}$}
	\begin{tabular}{r|rrrrr}
	$k$\textbackslash$s$ & 2 & 3 & 4 & 5 & 6 \\ \hline
	7	& 19 & 271 & 4098 & 57356 & 390370 \\
	8	& 23 & 302 & 5623 & 133385 & 2180416 \\
	9	& 20 & 313 & 6688 & 220255 & 6279611 \\ 
    10	& 19 & 320 & 7507 & 328959 & 13920691
	\end{tabular}
	\label{tab:fukumenDFA_ks}	
\end{minipage}
\hfill
\begin{minipage}[h]{.52\textwidth}
	\centering
  \caption{Construction times of $\widetilde{M}_{k,s}$ (sec.)}
	\begin{tabular}{r|rrrrr}
		$k$\textbackslash$s$ & 2 & 3 & 4 & 5 & 6 \\ \hline
	7	& $<0.01$ & $0.01$ & 0.6 & 10 & 90 sec. \\
	8	& $<0.01$ & 0.03 & 1.4 & 34 & 11 min. \\
	9	& $<0.01$ & 0.04 & 3.0 & 107 & 52 min.  \\ 
	10	& $<0.01$ & 0.07 & 6.1 & 307 & 210 min. 
	\end{tabular}
	\label{tab:fukumenDFA_ks_time}
\end{minipage}
\end{table}

\section{Analysis of Cryptarithms}
Cryptarithm automata $M_{k}$, $\widetilde{M}_{k}$ and $\widetilde{M}_{k,s}$ can be used as cryptarithm puzzle solvers as we have described in the previous section.
Moreover, they can be used as complete catalogues of solvable cryptarithms.
For example, one can count the number of base-$k$ solvable cryptarithms of size $n$ and
one can enumerate the base-$k$ solvable cryptarithm sequences by the length-lexicographic order.

\subsection{Counting Solvable Cryptarithms}
The number $F_{k}(n)$ of base-$k$ uniquely solvable cryptarithms of size $n$ is the number of the paths of length $n+1$ from the initial states to the accepting state $f_1$ in $\widetilde{M}_k$.
The number $G_{k}(n)$ of (not necessarily uniquely) solvable cryptarithms is obtained by adding the number of paths to the accepting state $f_2$ to this number.
Those numbers can be calculated by the standard technique using the adjacency matrix $A_k$ of the automaton in $\mrm{O}(m_k^3 \log n)$ time, where $m_k$ is the number of states of the automaton (i.e., $m_k$ is the number of rows (columns) of $A_k$).
Table~\ref{tb:num_fukumen} summarizes the numbers of uniquely and not necessarily uniquely solvable cryptarithms for $n \le 8$.
Note that there are no difficulties to compute $F_n(k)$ and $G_n(k)$ for bigger $n$.
Although we have computed $\widetilde{M}_7$, we were unable to calculate $F_7(n)$ and $G_7(n)$ even for small numbers $n$ by multiplying the adjacency matrices due to the size of the matrices.
Moreover, for $k=2,3$, we obtain $F_k(n)$ and $G_k(n)$ as explicit formulas of $n$ using Mathematica\texttrademark{} as follows.
\begin{gather*}
	F_2(n)=G_2(n) =  6 \times 4^{n-2}-3 \times 2^{n-2}
\\	F_3(n) = 4 \times 9^{n-1}-4 \times 5^{n-1}-3^{n-1}
\\	G_3(n) = 4 \times 9^{n-1}-2 \times 5^{n-1}-3^{n-1}
\end{gather*}
Unfortunately, Mathematica\texttrademark{} returned no answers for bigger $k \ge 4$ within 3 days on our environment.
\newcommand{\owr}[1]{\scalebox{0.87}[1]{\textbf{{\textcolor{black}{#1}}}}}
\newcommand{\owrm}[1]{\boldmath\scalebox{0.87}[1]{\textbf{\textcolor{black}{#1}}}}
\begin{table*}[h]
		\caption{The numbers $F_k(n)$ and $G_k(n)$ of uniquely and not necessarily uniquely solvable cryptarithms, respectively.
		 Among those, numbers shown with bold figures were not known in~\cite{ref:endo2013master}.
}
$F_k(n)$:\\
			\begin{tabular}{|c|r|r|r|r|r|r|r|r|r|r|}
				\hline
				 $k \backslash n$ & 1 & \phantom{00}2 & \phantom{}3 & \phantom{}4 & 5 & 6 & 7 & 8  \\ \hline
				2 & 0 & 3 & 18 & 84 & 360 & 1488 & 6048 & \owr{24384}  \\ \hline
				3 & 1 & 19 & 233 & 2443 & 23825 & 223939 & 2063993 & \owr{18821563}  \\ \hline
				4 & 1 & 46 & 1200 & 24094 & 431424 & 7326008 & 121032266 & \owr{1970599868}  \\ \hline
				5 & 0 & 42 & 3190 & 125940 & 3866438 & 106663574 & \owr{2797440502} & \owr{71604333066}  \\ \hline
				6 & 0 & 10 & 3470 & 336367 & 18978996 & \owr{847469530} & \owr{33983003374} & \owr{1292957034805} \\ \hline
			\end{tabular}  
$G_k(n)$:\\
      \begin{tabular}{|c|r|r|r|r|r|r|r|r|r|r|}
        \hline
        $k \backslash n$  & 1 & 2 & 3 & 4 & 5 & 6 & 7 & 8 \\ \hline
        2 & 0 & 3 & 18 & 84 & 360 & 1488 & 6048 & \owr{24384} \\ \hline
        3 & 1 & 23 & 265 & 2639 & 24913 & 229703 & 2093785 & \owr{18973439}  \\ \hline
        4 & 2 & 69 & 1463 & 26716 & 456639 & 7561377 & 123194460 & \owr{1990281467} \\ \hline
        5 & 2 & 115 & 4622 & 148483 & 4184478 & 110899540 & \owr{2852251360} & \owr{72299094358}  \\ \hline
        6 & 2 & 123 & 8650 & 498307 & 22931188 & \owr{933488391}  & \owr{35745728867} & \owr{1327783229135} \\ \hline
      \end{tabular}  
    \label{tb:num_fukumen}
\end{table*}

\subsection{Enumerating and Indexing Cryptarithms}
By depth-first search on a cryptarithm automaton, one can enumerate all the base-$k$ (uniquely) solvable cryptarithm sequences by length-lexicographic order.
Moreover, from an index number $i$, one can efficiently give the $i$\textsuperscript{th} (uniquely) solvable cryptarithm sequence. 
This can be computed in $\mrm{O}(m_k^3 n \log n)$ time, where $n$ is the length of the $i$\textsuperscript{th} cryptarithm, using powers of the adjacency matrix $A_k$.
Conversely, from a solvable cryptarithm of length $n$, the indexing number of it can be computed in $\mrm{O}(m_k^3 n \log n)$ time as well.
As examples, the first 30 ternary solvable cryptarithm sequences are given below.
\begin{center}
\scalebox{0.92}{
\begin{tabular}{rrrrr}
    {\tt aab\$\$\$}, & {\tt aaabbc\$\$\$}, & {\tt aab\$\$b\$\$\$}, & {\tt aab\$aa\$\$\$}, & {\tt aab\$ba\$\$\$},
\\  {\tt aab\$bb\$\$\$}, & {\tt aaba\$a\$\$\$}, & {\tt aabaab\$\$\$}, & {\tt aabb\$a\$\$\$}, & {\tt aabb\$b\$\$\$},
\\  {\tt aba\$aa\$\$\$}, & {\tt aba\$cc\$\$\$}, & {\tt abaaac\$\$\$}, & {\tt abacca\$\$\$}, & {\tt abbb\$b\$\$\$},
\\  {\tt abbbbc\$\$\$}, & {\tt abbc\$c\$\$\$}, & {\tt abbccb\$\$\$}, & {\tt abc\$\$a\$\$\$}, & {\tt abc\$\$b\$\$\$},
\\  {\tt abc\$ab\$\$\$}, & {\tt abc\$ba\$\$\$}, & {\tt abca\$b\$\$\$}, & {\tt abcb\$a\$\$\$}, & {\tt aaaaaabbc\$\$\$},
\\ {\tt aaaabbb\$b\$\$\$}, & {\tt aaaabbbbc\$\$\$}, & {\tt aaaabbc\$c\$\$\$}, & {\tt aaaabbccb\$\$\$}, & {\tt aaabab\$bb\$\$\$}\phantom{,}
\end{tabular}
}
\end{center}

\section{Conclusions and Discussions}
This paper proposed an algorithm to construct a DFA that accepts solvable cryptarithms under the base-$k$ numeral system.
Our construction method involves a technique to reduce the number of states more significantly than the classical minimization of DFAs by enriching transition edge labels.
We implemented the algorithm and constructed cryptarithm DFAs for $2 \leq k \leq 7$.
Moreover, by limiting the number of letters used in cryptarithms to $s \le k$, we managed to construct DFAs for even bigger bases.
Using those automata, we demonstrated that the numbers of base-$k$ solvable cryptarithms of $n$ digits are computable  for $2 \leq k \leq 6$.

Our compression technique is based on the symmetry among assignments.
Another type of symmetry is found between the first and second summand terms.
It is future work to take advantage of this type of symmetry to reduce the size of cryptarithm DFAs.
We are also interested in applying our DFAs for generating alphametics, which are cryptarithms with meaningful words.


\subsubsection*{Acknowledgments}
We thank to Kaizaburo Chubachi for assisting us in some of the experiments.
We also appreciate helpful comments by anonymous reviewers of CIAA~2018.
The work is supported in part by KAKENHI 15H05706 and 18K11153.

\bibliographystyle{plain}

\end{document}